\documentstyle[12pt]{article}
\begin{document}
\def\souligne#1{$\underline{\smash{\hbox{#1}}}$}
\def\square{\mathchoice\sqr34\sqr34\sqr{2.1}3\sqr{1.5}3}
\def\qed{ $ \quad\square $ }
\def\sigmab{\bar \sigma}
\def\EM{{\cal G}}
\def\Xb{\bar X}
\def\Ab{\bar A}
\def\Bb{\bar B}
\def\zb{\bar z}
\def\ab{\bar a}
\def\bb{\bar b}
\def\fb{\bar f}
\def\ib{\bar \imath}
\def\jb{\bar \jmath}
 \def\betab{\bar \beta}
\def\lb{\bar l}
\def\1b{\bar 1}
\def\nb{\bar n}
\def\ub{\bar u}
\def\Wb{\bar W}
\def\Cb{\bar C}
\def\xib{\bar \xi}
\def\mub{\bar \mu}
\def\nub{\bar \nu}
\def\phib{\bar \phi}
\def\bff{{\bf f}}
\def\bbff{\bar {\bf f}}
\def\bfe{{\bf e}}
\def\bbfe{\bar {\bf e}}
\def\bfX{{\bf X}}
\def\bfY{{\bf Y}}
\def\partialb{\bar \partial}
\def\au{\underline a}
\def\bu{\underline b}
\def\Au{\underline A}
\def\Bu{\underline  B}
\begin{titlepage}
\nopagebreak
\begin{flushright}

LPTENS--91/29\\
                September~1991
\end{flushright}

\vfill
\begin{center}
{\large\bf
W-GEOMETRIES}

\vfill
{\bf Jean-Loup~GERVAIS}\\
and \\
{\bf Yutaka~MATSUO}\footnote{present address
Niels Bohr Institute Blegdamsvej 17 DK-2100 Copenhagen $\emptyset$.} \\
\bigskip
Laboratoire de Physique Th\'eorique\\
Ecole Normale Sup\'erieure\footnote{Unit\'e propre du
Centre National de la Recherche Scientifique,
associ\'ee \`a l'Ecole Normale Sup\'erieure et \`a l'Universit\'e
de Paris-Sud.},\\
24 rue Lhomond, 75231 Paris CEDEX 05, ~France.\\
{\footnotesize \LaTeX \ file available from
hepth@xxx.Lanl.GOV}
\end{center}
\vfill

\begin{abstract}
\noindent
It is shown that, classically, the  W-algebras are directly related to the
extrinsic  geometry of the embedding of two-dimensional
manifolds with chiral parametrisation (W-surfaces) into
higher dimensional
K\"ahler manifolds. We  study the local and the global
geometries of such embeddings, and connect them to Toda
equations. The additional variables  of
the related KP hierarchy are shown to  yield a specific
coordinate system of the target-manifold, and this allows us
to prove  that W-transformations are simply particular
diffeomorphisms of this  target space. The  W-surfaces
are  shown to be
instantons of  the corresponding non-linear $\sigma$-models.

\end{abstract}
\vfill

\end{titlepage}

\section{Introduction                }
In many ways, W-algebras are natural generalisations of the Virasoro
algebra. They were first introduced as  consistent operator-algebras
 involving
operators of spins higher than two\cite{Z}. Moreover, the Virasoro
algebra is intrinsically related with the Liouville  theory which is
the Toda theory associated with the Lie algebra $A_1$, and this
relationship extends to W-algebras which  are in correspondence with
the family of conformal Toda systems associated with arbitrary simple
Lie algebras\cite{BG1}. Another point is that the deep connection between
Virasoro algebra and KdV hierarchy has  a natural extention\cite{GY}
  to  W-algebras
and higher KdV (KP) hierachies\cite{Sato}\cite{DS}.

On the other hand, W-symmetries exhibit strikingly novel features. First,
they  are basically non-linear algebras. Since the transformation
laws  of primary fields  contain higher derivatives,  product of primaries
 are not primaries at the classical level. Naive tensor-products of
commuting representations do not form representations. A related novel
feature is that W-algebras generalise the diffeomorphisms of the circle
by including derivatives of degree higher than one. Going beyond linear
approximation (tangent space ) is  a highly non-trivial step.
Taking higher order derivatives changes the shape of the world-sheet
in the target-space, thus W-geometry should be related to the extrinsic
geometry of the embedding.
Finally, Virasoro algebras are notoriously related to Riemann surfaces.
The W-generalisation of the latter notion is a fascinating problem.

In the present letter, we show that 2D surfaces  $\Sigma$ with
 chiral parametrisations,
 embedded into   K\"ahler manifolds,    are the geometrical
framework where
the features just recalled are naturally realised, at
the classical level. This idea which
hopefully will become clear below was supported at the beginning
by several hints. First,  the extrinsic geometry of abstract Riemann surfaces
was beautifuly discussed\cite{GH},  using the   complex analogue of
Frenet's  formalism,  by lifting the curve to Grassmannian manifolds
with  the Fubini-Study metric which is  a K\"ahler metric associated
with the standard  metric
 on $CP^n$. Second, one of us has just shown\cite{M} that the
 free fermion formalism of the KP hierarchy, and its related Grassmannian
 manifold --- which is directly related to   the one just mentionned ---
has a close connection with W-algebras, the additional variables
of the KP systems, allowing to linearise the W-transformations.

This letter briefly summarizes  our results. Details will be given
elsewhere.
In section 2 we study the local structure of the
embedding at generic regular points
where Taylor expansion gives a local frame. Section 3 is concerned with
singular points and with the
   global
Pl\"ucker formula that describes the global topology of W-surfaces.
Section 4 discusses  the search for the  action principle that would
underly  the
picture just derived. We show that the W-surfaces just introduced
are  instantons
of the  non-linear $\sigma$-models with the same K\"ahler
potential.

\section{Local structure of the embedding at regular points}
Our basic strategy is to study embeddings of two-dimensional surfaces
$\Sigma$ given by equations of the form
\begin{equation}
\label{1}
X^A=f^A(z), \, A=1,\, \cdots,\, n, \quad
\Xb^{\Ab}=\fb^{\Ab}(\zb), \, \Ab=1,\, \cdots,\, n,
\end{equation}
 in a 2n-dimensional target manifold $\cal M$ with a  K\"ahler
metric\footnote{In this first general part, the K\"ahler condition
is not really neded. The same discussion applies to a general
Hermitean manifold. The K\"ahler condition will play an essential r\^ole
in the free-fermion approach, and in sections  3 and 4.}
\begin{equation}
\label{2}
G_{A\Bb}=G_{\Bb A}=\partial_A \partial_{\Bb} {\cal K}(X, \Xb).
\end{equation}
One may either work with  Minkowski coordinates where $z=\sigma+\tau$,  and
$\zb =\sigma-\tau$,  are light-cone coordinates, or perform a Wick rotation
and work with the complex $z$ variable (isothermal coordinates).
The first description makes it clear
that $f^A$ and $\fb^{\Ab} $ are independent functions,
so that we are not dealing
with a single complex curve as is often done in this context.
In the present section, we shall be concerned with regular
points of $\Sigma$ where Taylor expansion gives a local frame, so
that a discussion of the Frenet type applies.

Our first result is that  Toda fields naturally
arise
from the   Gauss-Codazzi
equations of this embedding.
 These   equations  are
 integrability conditions for derivatives of
the tangent and   of the normals
to  the surface. The latter are introduced
by extending
Frenet-Serre formula as follows. Define
\begin{equation}
\label{3}
g_{i \jb}\equiv G_{\! A\Bb} (f,\fb) \>\>\partial^{(i)} \! f^A(z)\>
\partialb^{(\jb)}\!  \fb^{\Ab}(\zb),
\end{equation}
\begin{equation}
\label{4}
\Delta_l\equiv \left | \begin{array}{ccc}
g_{1 \1b} & \cdots  & g_{l \1b} \\
\vdots    &         & \vdots  \\
g_{l \1b} & \cdots  & g_{l \1b}
\end{array}
\right | \equiv e^{-\phi_l}.
\end{equation}
$\partial $ and $\partialb$ are short hands for
$\partial/\partial z$ and $\partial /\partial \zb$ respectively.
$\partial^{(i)}$ stands for $(\partial)^i$.  Vectors in $\cal M$ are denoted
by boldface letters. It is straightforward to check that a set
of $2n$ orthonormal vectors
   is given by
 (upper indices in between parenthesis denote
derivatives)
\begin{equation}
\label{5}
{\bf e}_a = {1\over \sqrt{\Delta_a \Delta_{a-1}}}
\left | \begin{array}{ccc}
g_{1 \1b} & \cdots  & g_{a \1b} \\
\vdots    &         & \vdots  \\
g_{1 {\overline {a-1}} } & \cdots  & g_{a {\overline {a-1}}} \\
{\bff^{(1)}} & \cdots & {\bff^{(a)}}
\end{array}
\right |, \quad a=1, \cdots, n,
\end{equation}
\begin{equation}
\label{6}
\bbfe_a = {1\over \sqrt{\Delta_a \Delta_{a-1}}}
\left | \begin{array}{ccc}
g_{\1b 1} & \cdots  & g_{\1b a} \\
\vdots    &         & \vdots  \\
g_{\1b a-1} & \cdots  & g_{{\overline {a}} a-1} \\
\bbff^{(1)} & \cdots & \bbff^{(a)}
\end{array}
\right |, \quad a=1, \cdots, n,
\end{equation}
where  determinants are to be computed for each
component of the last line.
Indeed, if we denote by $( \bfX, \bfY)$ the inner  product
$G_{A \Bb} (X^A Y^{\Bb} +Y^A X^{\Bb})$ in $\cal M$, we have
$( \bfe_a, \bfe_b)=( \bbfe_a, \bbfe_b)=0$,
$( \bfe_a, \bbfe_b) =\delta_{a,b}$. It will be convenient at some
point to treat bar and unbar indices together. This is done by using
underlined letters.  In this way, $  \bfe_a$ and $\bbfe_a$ are denoted
collectively  as $\bfe_{\au}$. One has
\begin{equation}
\label{7}
( \bfe_{\au}, \bfe_{\bu})=\eta_{\au\, \bu},  \quad
\eta_{a \bb} =\delta_{a,\bb},\quad  \eta_{a\, b} =\eta_{\ab\,  \bb}=0
\end{equation}
${\bfe}_1$ and  ${\bbfe}_1$ are tangents to the surface,
while the other vectors are clearly normals.
One may show that the  ${\bfe}_a $'s  satisfy  equations  of the form
\begin{equation}
\label{8}
\nabla \bfe_a=\omega_{z a}^{\>b} \bfe_b, \quad
\partialb \bfe_a=\omega_{\zb a}^{\>b} \bfe_b,
\end{equation}
\begin{eqnarray}
\label{9}
\omega_z =-{1\over 2} \sum_{i=1}^{n} H_i
\left (\partial\phi_i -\partial\phi_{i-1}\right)
+\sum_{i=1}^{n-1} \exp \left ( \sum_{j=1}^{n} K_{i j} \phi_j/2 \right )
E_{-i}\nonumber \\
\omega_{\zb} ={1\over 2} \sum_{i=1}^{n} H_i
\left (\partialb\phi_i -\partialb\phi_{i-1}\right)
-\sum_{i=1}^{n-1} \exp \left ( \sum_{j=1}^{n} K_{i j} \phi_j/2 \right )
E_{i},
\end{eqnarray}
where $\phi_0\equiv0$.
 $\nabla $ stands for $f^{(1)A}\nabla_A$,   $\nabla_A$ being  the
target-space covariant derivative, and
 $K$ is the Cartan matrix of $A_{n}$.  $H_i$, $E_{\pm i}$ are
infinitesimal generators of $U_n$ in the Chevalley basis, for
the fundamental representation.
  $H_i$ generate
the Cartan subalgebra, and $E_{\pm i}$ are associated with  the usual  set
of primitive roots of $A_n$.  Remarkably, one sees that the right member
is just the Toda Lax pair\cite{LS}. Toda equations
are equivalent to the zero-curvature condition on $\omega$.  In
the language of Riemannian
geometry\cite{E},
$\omega_{z 1}^{\>b}$ and $\omega_{z a}^{\>b}$ $a>1$
 give  the second  and third fundamental forms respectively.
  $\bbfe_a$ satisfy similar equations.
The corresponding
Toda Lax pair is in the adjoint of the fundamental representation\footnote{
A connection between Toda Lax pairs and Gauss-Codazzi equations already
appears in ref.\cite{S}. There, however, the former is defined
 in an abstract
space whose geometrical meaning seems unclear.}.

Going to the
second derivatives, one deduces from the above that
\begin{eqnarray}
\label{12}
\bigl [ \nabla, \partialb \bigr ] \bfe_a& = &
F_{z\zb a}^{\> \> b} \bfe_b   \nonumber \\
F_{z\zb }&=&
\sum_{i=1}^{n} H_i \partial \partialb\left (  \phi_i-\phi_{i-1}\right )+
\sum_{i=1}^{n-1} (H_i-H_{i+1})
\exp \left ( \sum_{j=1}^{n} K_{i j} \phi_j \right )
\end{eqnarray}
On the other hand,
$\bigl [ \nabla, \partialb \bigr ] \bfe_a=
{\cal R} (\bff^{(1)}, \bbff^{(1)})_a^b\bfe_b$ where
${\cal R} $ is the target-space curvature tensor. Thus we get
\begin{equation}
\label{13}
F_{z\zb a}^{\> \> b} \bfe_b=
{\cal R} (\bff^{(1)}, \bbff^{(1)})_a^b\bfe_b,
\end{equation}
which are the Gauss-Codazzi equations\cite{E} for the embedding
considered\footnote{Ideas,  that are somewhat related   to ours,
  have already been put
forward in ref.\cite{SS}.}.
Geometrically, the vectors
$\bfe_{\au}$  span a moving frame, where the metric is constant
and equal to $\eta_{\au\, \bu}$. Thus the $\bfe_{\au}$ are 2n-beins,
  $\omega_z$, $\omega_{\zb}$ are the two components of the spin
connection $\omega_{\Au}$ along the surface,  and the meaning of
Eq. \ref{13}
is clear.

Let us now turn to W-transformations. They are defined, as usual, to be
of the form $\delta_W f^{A}={\cal D}_W f^{A}$ where ${\cal D}_W$ is a
differential
operator in $z$. Higher derivatives may be eliminated using the
fact that the  functions $f^{A}$  are automatically
solutions of the following
 differential
equation
\begin{equation}
\label{W1}
\left | \begin{array}{cccc}
f^1 & \cdots  & f^n & f \\
\vdots    &   & \vdots      & \vdots  \\
f^{(n)\, 1}  & \cdots  & f^{(n)\, n}  & f^{(n)}
\end{array}
\right |
\equiv \left(\sum_{k=0}^{n} U_{n-k}\partial^{(k)}\right ) f=0.
\end{equation}
This allows us first to rewrite the W-transformation as
$\delta_W f=\sum_{a=1}^n w_a f^{(a)}$. Moreover, taking the derivative of
Eq.\ref{W1}, one derives an equation of the form
$f^{(n+1)}=\sum_{a=1}^n \lambda_a f^{(a)}$. Then
one easily verifies that the transformation laws  of $f^{(a)}$ may be
put under the  matrix form
\begin{eqnarray}
\label{W2}
\delta_W f^{(a)}=\sum_{b=1}^n B_b^a f^{(b)},
\>
&B_b^a&= \sum_{d=1}^n\left ( (\partial +q+\lambda)^{a+1}\right )_b^d w_d
\nonumber\\
q=\left (\begin{array}{ccccc}
0 & 0 & 0 &\cdots & 0 \\
1 & 0 & 0 &\cdots & 0 \\
0 & 1 & 0 &\cdots & 0 \\
\vdots & \ddots & \ddots &\,  & \vdots \\
0 & \,  & 0 &1 & 0 \end{array}
\right ), \>
&\lambda&=\left (\begin{array}{ccccc}
0 & 0 & 0 &\cdots & \lambda_1 \\
0 & 0 & 0 &\cdots & \lambda_2\\
\vdots & \, & \ddots &\,  & \vdots \\
0 & \, & \cdots &0  & \lambda_{n-1}\\
0 & 0  & \cdots &0 &  \lambda_n
 \end{array}
\right ).
\end{eqnarray}
 According to Eqs. \ref{5}, \ref{6}, this leads to a matrix transformation
for the $\bfe_{\au}$ which is equivalent to a gauge transformation
of  $\omega_z$ and $\omega_{\zb}$. By construction it
is  such that Eqs. \ref{9}, and
\ref{12} keep the same form. Thus it is the same as  a
W-transformation\footnote{Note that these are defined in
\cite{BG1}, through  canonical
Poisson brackets that  do not assume that Toda equations  hold.}
of $\phi_l$ in Toda theory\cite{BG1}.
Geometrically,  and since
 the $\bfe_{\au}$  are 2n-beins,  W-transformations
appear as  particular   transformations of the local Lorentz group,
 which leave the
form of the Toda Lax pair invariant. An important point is that
W-transformations in general mix the tangent  and the normals at
a given point of $\Sigma$.
 Thus they move the surface     and the
distinction between intrinsic and extrinsic geometry is
not W-invariant.

Next we show that the geometry of
 W-transformations   becomes  transparent  if we
 introduce   the additional variables
 $z^k$, $\zb^k$, $k=2$, $\cdots$ $n$,  that
play the r\^ole of the additional times of the KdV hierarchy\cite{Sato}.
This was already shown in  the free-fermion approach of \cite{M}.
Inspired by this analogy we introduce these additional variables  by
extending $f^A$ (or  $\fb^{\Ab}$) to functions of   $z$, $ z^2$,
$\cdots$, $z^n$ (or $\zb$, $ \zb^2$, $\cdots$, $\zb^n$)
 that satisfy the differential
equations
\begin{eqnarray}
\label{14}
\left ({\partial\over \partial z^k}-
{\partial^k\over(\partial z)^k}\right )
f^A(z, z^2, \cdots, z^n)= 0, \quad
f^A(z, 0, \cdots, 0)=f^A(z), \nonumber \\
\left ({\partial\over \partial\zb^k}-
{\partial^k\over(\partial \zb)^k}\right )
\fb^{\Ab}(\zb, \zb^2, \cdots, \zb^n)=0, \quad
\fb^{\Ab}(\zb, 0, \cdots, 0) =\fb^{\Ab}(\zb). \nonumber \\
\end{eqnarray}
With these  additional variables, the W-transformations
may be rewritten as
(from now on $z$ and $\zb$
are replaced by $z^1$, and $\zb^1$ if needed for compactness)
\begin{equation}
\label{16}
\delta_W f^{A}\equiv
\sum_{a=1}^n w^a\partial^{(a)} f^{A}= \sum_{a=1}^n  w^a \partial_a f^{A}
\end{equation}
where $\partial_a\equiv \partial/\partial z_a$. By using the
differential equations \ref{14}, one extends this last
relation to the whole manifold $\cal M$.  Then the W-transformations
become
$\delta_W f^{A}\sim f^{A}(z+w^1, z^2+w^2,\cdots, z^n+w^n)-
f^{A}(z, z^2,\cdots, z^n)$. They thus
   reduce to  changes  of  variables
\begin{equation}
\label{17}
\delta_W z^k=w^k(z,\,z^2,\,\cdots,\, z^n).
\end{equation}
  Eq.\ref{3}  may now  be rewritten as
$g_{i \jb}\equiv G_{\! A\Bb}(f,\fb) \>\partial_i f^A\>
\partialb_{\jb} \fb^{\Ab}$,
which simply  means that
$g_{i \jb} dz^i d\zb^{\jb}=G_{A\Bb}dX^A d\Xb^{\Bb}$. Thus one sees that
 the variables
$z^k$, and  $ \zb^k$ are  a set of coordinates on $\cal M$ such that
\footnote{Of course there may be global obstructions, so that they
only cover the component of $\cal M$ which is connected to the surface.}
the embedded surface has equations  $z^k=0$, $ \zb^k=0$ for
$k\not=1$. Thus   {\bf W-transformations are
particular reparametrisations
of the target space $\cal M $}. Concerning the moving frame,  and
generalising the above discussion, one defines 2n-beins $\bfe_{\au}$
on $\cal M$, by using the additional coordinates.
  Eqs.  \ref{8}-\ref{13} are now a  part of
 the standard equations
relating the spin connection $\omega_{\Au\,  \au}^{\> \bu}$, its curvature
$F_{\Au\,  \Bu}$  and the metric tensor $\cal R$. From this
viewpoint,
 the W-transformations appear as
local Lorentz transformations  that leave the form of
Eq.\ref{9} invariant on $\Sigma$.

As it is evident from Eqs. \ref{4},  \ref{5}, \ref{6},
the nature of the chiral embedding is
deeply connected with free fermions.
Indeed, when the target space is simple, we can get neat expressions
for Eqs.  \ref{4}, \ref{5}, \ref{6}   as
expectation values between quantum states of     non-relativistic
fermions.
This is the non-chiral version of the method
developed in \cite{M} and we shall use the same notation.

The simplest situation is when the target space is a flat space,
i.e. $ C^n$.  The key observation is that the metric
$g_{i\jb}$ can be written as expectation values  between
one-fermion states,
\begin{eqnarray}
g_{i\jb}& =& <\emptyset|\psi_{\jb-1}^*{\cal G}^{(1)}_{C^n}
\psi_{i-1}|\emptyset>,\nonumber \\
{\cal G}^{(1)}_{C^n}& = &\sum_{A=1}^n e^{\bar{J}_1
\bar{z}}\psi_{\textstyle\partialb\bar{f}^A}|\emptyset>
<\emptyset|\psi^*_{\textstyle\partial f^A}\> e^{J_1 z}.
\end{eqnarray}
Once one recognizes this fact,  Wick's theorem tells us that the
determinant Eq.\ref{4} is given by the expectation value
in the  $a$-particle state,
$\Delta_a = <a|{\cal G}^{(a)}_{C^n}|a>,
$
where
\begin{equation}
{\cal G}^{(a)}_{C^n} = \sum_{A_1<\cdots<A_a}
e^{\bar{J_1}\bar{z}}\psi_{\textstyle\partialb \bar{f}^{A_1}}
\cdots\psi_{\textstyle\partialb \bar{f}^{A_a}}|\emptyset>
<\emptyset|\psi^*_{\textstyle\partial f^{A_a}}\cdots
\psi^*_{\textstyle\partial f^{A_1}}\> e^{J_1 z}.
\end{equation}
In this language, the orthogonality condition \ref{7} or the extended
Frenet formula \ref{8} is the direct consequence of  Hirota's
equations,
\begin{equation}
\sum_{i=0}^\infty \psi^*_i{\cal G}^{(\ell)} \otimes
\psi_i{\cal G}^{(s)}
= \sum_{i=0}^\infty {\cal G}^{(\ell-1)}\psi^*_i \otimes
{\cal G}^{(s+1)}\psi_i.
\end{equation}
( This is the re-interpretation of  Hirota's equations\cite{TU}
in the free-fermion language).  On the other hand, one is in
a much more interesting situation  when the target space
has a
non-trivial topology.  The first such  example is $ CP^n$.
Introduce the inhomogeneous coordinates $[X^0\equiv 1, X^1,\cdots, X^n]$
and the Fubini-Study metric, $G_{A\bar{B}} =
\partial_A\bar{\partial}_{\Bb}
\ln \sum_{C=0}^n X^C\bar{X}^{C}$.  The induced world-sheet K\"ahler
potential is given by,
\begin{equation}
{\cal K}_{ CP^n} = \ln \tau_1 \equiv
\ln \sum_{A=0}^n f^A(z)\bar{f}^A(\bar{z}).
\end{equation}
We should first remark that the $\tau$-function can be written as
an expectation value  $\tau_1 = <1|{\cal G}^{(1)}_{CP^n}|1>$.
with
\begin{equation}
{\cal G}^{(1)}_{ CP^n} = \sum_{A=0}^ne^{\bar{ J}_1\bar{z}}
\psi_{\textstyle\bar{f}^A}|\emptyset>
<\emptyset|\psi^*_{\textstyle f^A}\>e^{J_1z}.
\end{equation}
The corresponding state  is a one-particle state, in contrast
with the previous case.
Then it is easy to check that the metric $g_{i\jb}$ can be written
as the {\it two}-particle expectation value,
\begin{equation}
g_{i\jb} = \frac{1}{\tau^2_1}(\tau_1\partial^{(i)}\partialb^{(\jb)}\tau_1-
\partial^{(i)}\tau_1\partialb^{(\jb)}\tau_1) =
\frac{1}{\tau_1^2}<1|\psi^*_{\jb}\EM^{(2)}_{ CP^n}\psi_i|1>.
\end{equation}
Using  again Wick's theorem,
the determinant Eq. \ref{4} can be expressed as a fermion
expectation value:
$\Delta_\ell = \tau_{\ell+1}/\tau_1^{\ell+1}$,
$\tau_{\ell} = <\ell|\EM^{(\ell)}_{ CP^n}|\ell>$.
Frenet's theorem follows again from Hirota's  equations.

In $C^n$ and $CP^n$, the target-space has constant sectional
curvature. Toda equations follow from Eq.\ref{13}.  This fact
again illustrates the
essential link between the embedding problem and Toda equations.
 One can, in
addition,  make contact with the conformally reduced WZNW theory\cite{WZW}.
This will be explained somewhere else.

For $C^n$ and $CP^n$, the K\"ahler potentials take the same form
but the matrix element is between zero- and one- particle states
respectively.
More generally, when the target space is the Grassmannian manifold
$Gr_{n+k,k}$, the K\"ahler potential can be related to the
$k$-particle fermion expectation value.
For  these simple target-spaces, the
 change of the K\"ahler potential can be reduced to
changes  of the vacuum and of the operator  $\EM$.
It natural to conjecture that
there exists  similar free-fermion expressions  for the embedding in any
K\"ahler manifold.  It will be very important to spell out
this correspondence because  the free-fermion formalism
is a very  natural tool to understand the W-geometries.
For example, as it was observed previously, the
dependence on the higher coordinates can be  easily
introduced by the simple change of the Hamiltonian,
$\exp(J_1z)\rightarrow \exp(\sum_{s=1}^n J_s z^s)$.
Furthermore, the W-transformations of the $\tau$-functions
can be written linearly through the bosonization of this
free fermion.

For $n\to \infty$, very interesting phenomena may take place.
This fermion formalism, with a non-trivial vacuum
structure   becomes  closely related to the
fermions of matrix-models.

\section{Global structure of the embedding}
The Grassmannian manifold $G_{n+k,\, k}$ is the set of
$(n+k)\times k$ matrices $W$ with the equivalence
relation $W\sim aW$, where $a$ is a $k\times  k$ matrix.
A chiral embedding in  $G_{n+k,\, k}$  is thus defined by
\begin{equation}
\label{G.1}
W(z)=
\left ( \begin{array}{ccc}
f^{1, 1}(z) & \cdots  & f^{n+k,\, 1}(z) \\
\vdots    &         & \vdots  \\
f^{1, k}(z) & \cdots & f^{n+k,\, k}(z)
\end{array}
\right )
\end{equation}
Following Ref.\cite{KN}, we introduce the K\"ahler
potential ${\cal K}_k\equiv \ln (\det W W^T)$, that is,
\begin{eqnarray}
\label{G.2}
e^{{\cal K}_k}&=& \nonumber \\
\sum_{1\leq i_1<\cdots <i_k\leq n+k}
& &\left |  \begin{array}{ccc}
f^{i_1, 1}(z) & \cdots  & f^{i_k,\, 1}(z) \\
\vdots    &         & \vdots  \\
f^{i_1, k}(z) & \cdots & f^{i_k,\, k}(z)
\end{array}
\right |
\left | \begin{array}{ccc}
\fb^{i_1, 1}(\zb) & \cdots  & \fb^{i_k,\, 1}(\zb) \\
\vdots    &         & \vdots  \\
\fb^{i_1, k}(\zb) & \cdots & \fb^{i_k,\, k}(\zb)
\end{array}
\right |
\end{eqnarray}
The difference is that,  in our case, $\fb$ will  not the
complex conjugate of $f$.
{}From the embedding $\Sigma\to CP^n$, we can canonically construct
the kth associated embedding $\Sigma\to Gr_{n+1,\, k+1} \hookrightarrow
P(\Lambda^{k+1} C^n)$.   We  let $f^{l,s}=
\partial^{(s-1)} f^l(z)$ in Eq.\ref{G.1}, for $s=1$, $\cdots$, $k+1$,
$l=0$, $\cdots$, $n$. The induced metric on the
corresponding Riemann surface is simply
\begin{equation}
\label{G.3}
g_{z\, \zb}^{(k)}=\partial \partialb \ln \tau_{k+1}
\end{equation}
so that the Toda field appears naturally.
The infinitesimal Pl\"ucker formula is derived by
computing  the curvature
\begin{equation}
\label{G.4}
R_{z\, \zb}^{(k)}\equiv -\partial \partialb \ln  g_{z\, \zb}^{(k)}
=-\partial \partialb \ln \left ( {\tau_{k+2} \tau_{k}\over \tau_{k+1}^2}
\right ),
\end{equation}
and this gives
\begin{equation}
\label{G.5}
R_{z\, \zb}^{(k)}=-g_{z\, \zb}^{(k+1)}+2g_{z\, \zb}^{(k)}
-g_{z\, \zb}^{(k-1)}.
\end{equation}
The global  Pl\"ucker formula contains the k-th instanton number
\begin{equation}
\label{G.6}
d_k\equiv {i\over 2\pi} \int _{\Sigma} dzd\zb g_{z\, \zb}^{(k)},
\end{equation}
and follows from Gauss-Bonnet theorem for $g_{z\, \zb}^{(k)}$:
\begin{equation}
\label{G.7}
{i\over 2\pi} \int _{\Sigma} dzd\zb R_{z\, \zb}^{(k)}
=2-2g+\beta_k.
\end{equation}
$\beta_k$ is defined as a sum of the   k-th ramification indices.
Near a singular point where there is an obstruction to the construction
of the moving frame, the behaviour is of the form
\begin{equation}
\label{G.8}
g_{z\, \zb}^{(k)}\sim (z-z_0)^{\beta_k(z_0)} (\zb-\zb_0)^{\betab_k(\zb_0)}
\tilde g_{z\, \zb}^{(k)},
\end{equation}
where $\tilde g_{z\, \zb}^{(k)}$ is regular at $z_0, \zb_0$. Since we do not
assume that $\overline {f(z)}=\fb(\zb)$,  $\beta_k(z_0)$ and
$\betab_k(\zb_0)$ may be different.  $\beta_k$ is defined by
\begin{equation}
\label{G.9}
\beta_k\equiv {1\over 2}
\sum_{ (z_0, \zb_0)\in \Sigma}\left ( \beta_k(z_0)+
\betab_k(\zb_0)\right ).
\end{equation}
Finally we arrive at the global Pl\"ucker formula

\begin{equation}
\label{G.10}
2-2g+\beta_k=2d_k-d_{k+1}-d_{k-1}, \quad \left \vert
\begin{array}{cc} k&=0,\cdots, n-1\nonumber \\
d_n&\equiv 0, \>
d_{-1}\equiv 0 \end{array}\right.
\end{equation}
Using this formula, we find that there are n independent topological numbers
($d_0$, $\cdots$, $d_{n-1}$), which characterize the global topology of
W-surfaces. A direct consequence of this observation is that    W$_{n+1}$-
strings  have n coupling constants which play the same r\^ole
as  the genus for  the usual
string theories. Eq. \ref{G.10} may be understood as the index theorem
for W-surfaces.

\section{Searching for an action principle}
In the preceeding sections, we have exhibited the  geometry
of W-surfaces. They are characterized by their  chiral parametrisation
$X^A=f^A(z)$, $\Xb^{\Ab}=\fb^{\Ab}(\zb)$.
This is of course a very restricted  set since a generic
   2D manifold embedded in $\cal M$   has equations
 $X^A=u^A(z,\zb)$, $\Xb^{\Ab}=\ub^{\Ab}(z,\zb)$.
For W-surfaces,
$u^A$ and $\ub^{\Ab}$ satisfy the Cauchy-Riemann  equations
which are self-duality equations.
Thus \souligne{W-surfaces are instantons in $\cal M$}.
At this point one should remember the exciting developments of the
seventies concerning instantons  in $CP^n$ models\cite{CPN}, and more
generally in  K\"ahler manifolds\cite{P}.
Instanton solutions  minimize  the action of  the associated
non-linear $\sigma$-model\cite{P}. Let us recall briefly the general
argument\cite{P} for completeness.  Consider a general K\"ahler manifold $M$
with coordinates $\xi^\mu$ and $\xib^{\mub}$ and metric $h_{\mu \mub}$.
In the  non-linear $\sigma$-model, the action associated with
an arbitrary
2D manifold of $M$   with equations $\xi^\mu=\phi^\mu(z,\zb)$,
$\xib^{\mub}=\phib^{\mub}(z,\zb)$ is given by
\begin{equation}
\label{I.1}
S={1\over 2} \int d^2x\> h_{\mu \mub}\,  \partial_j \phi^\mu
\partial_j \phib^{\mub}.
\end{equation}
In this part, we let $z=x_1+ix_2$, $\partial_j=\partial/\partial
x_j$.  It is easy to see that
\begin{equation}
\label{I.2}
S\geq cQ,\quad Q= {i\over 2c} \int d^2x\> \epsilon_{jk}\>
h_{\mu \mub} \partial_j \phi^\mu
\partial_k \phib^{\mub}.
\end{equation}
($c$ is a suitably chosen constant)
Clearly the equality is achieved if $\partial_j\phi^\mu=\pm i
\epsilon_{jk}  \partial_k\phi^\mu$.  These solutions are
the standard instantons
of K\"ahler manifolds.

At this point a general remark is in order. In the problems considered
earlier\cite{CPN}\cite{P}, the natural reality condition is
$\left ( \phi^\mu(z)\right )^*=\phib^{\mub}(z^*)$. Such is not the
case for W-surfaces as we show next, since left- and right- moving
modes  are
not correlated (apart from the zero-modes)  for conformal systems
without boundaries.
The physical requirement  is that
the Toda fields be real in Minkowski space. Thus
$g_{i \jb}\equiv G_{\! A\Bb} (f,\fb) \>\>\partial^{(i)} \! f^A(z)\>
\partialb^{(\jb)}\!  \fb^{\Ab}(\zb)$ must be real for $z=\sigma+\tau$,
and $\zb=\sigma-\tau$,  real. This is achieved by  conditions of the form
\begin{equation}
\label{I.3}
\left (f^A(z)\right )^*= C^A_B f^A(z^*),\quad
\left (\fb^{\Ab}(\zb)\right )^*= \Cb^A_B
\fb^{\Ab}(\zb^*),
\end{equation}
where the conjugation matrices $C$ $\Cb$  must be such that
\begin{equation}
\label{I.4}
(G_{\! A\Ab})^* C^A_B \Cb^{\Ab}_{\Bb}=
G_{\! B\Bb}.
\end{equation}
With the usual reality condition the  self-duality equations for
$\phi^\mu$ and $\phib^{\mub}$ are equivalent. This is no more
true
in our problem. Note that the lower bound $S=Q$ is reached as
soon as
the self duality equation holds for $\phi$ or $\phib$. Thus there seems
to exist more general instantons than the W-surfaces  we have
considered so far. We leave this problem for further studies.

Going back to our main line, we note that the analysis just recalled
may be carried out  not only on $\cal M$ but also on the Grassmannian
manifolds whose construction was recalled  in section 3 for $CP^n$.
For each $k$, Eq.\ref{G.3}
precisely means that the action is tological, and that the lower bound
$S=Q$ is reached. Considering these higher Grassmannians  may
shed new lights
on the instanton problems of the seventies.

This non-linear $\sigma$-model action is quite interesting but is not
the final answer. Since W-symmetries are  so important,
one should introduce
a W-invariant action. We have seen that the W-symmetry is a part of
general covariance so that one should  start from an action
invariant under
diffeomorphisms of the target space.  On the other hand, W-transformations
applied to a W-surface, do not  generically  leave  it invariant
as a geometrical
object. Thus the action must be topological,  since
it does not depend upon the shape of the W-surface. Moreover,
since
the position of the W-surface is irrelevant, there is no real
distinction
between intrinsic and extrinsic geometry. This is why the extrinsic
curvature-components become part of the conformal theory on the
world-sheet. A possible candidate for the W-action is a  topological
Yang-Mills action
in 2n dimensions, involving  the spin connection $\omega$.
We leave this exciting problem for
further investigations.
\bigskip

\noindent {\bf Acknowledgements}
One of us (J.-L. Gervais) is grateful to M. Saveliev for stimulating
conversations and for pointing out his inspiring article\cite{S}.

\end{document}